\documentclass[twocolumn,showpacs,amssymb,aps,prl,floatfix]{revtex4-1}

\usepackage{epsfig}

\newcommand{\be}{\begin{equation}}
\newcommand{\ee}{\end{equation}}
\newcommand{\beq}{\begin{equation}}
\newcommand{\eeq}{\end{equation}}
\newcommand{\bea}{\begin{eqnarray}}
\newcommand{\eea}{\end{eqnarray}}

\newcommand{\nn}{\nonumber}
\newcommand{\e}{\hbox{e}\,}
\newcommand{\tr}{\hbox{Tr}}

\newcommand{\xv}{{\mathbf x}}
\newcommand{\bra}{\langle}
\newcommand{\ket}{\rangle}

\begin{document}

\title{Simulating QCD at nonzero baryon density to all orders in the hopping parameter expansion}

\author{Gert Aarts$^a$}
\author{Erhard Seiler$^b$}
\author{D\'enes Sexty$^c$}
\author{Ion-Olimpiu Stamatescu$^c$}

\affiliation{$^{a}$Department of Physics, College of Science, Swansea University,
Swansea SA2 8PP, United Kingdom}
\affiliation{$^{b}$Max-Planck-Institut f\"ur Physik (Werner-Heisenberg-Institut)
M\"unchen, Germany}
\affiliation{Institut f\"ur Theoretische Physik, Universit\"at Heidelberg, Heidelberg, Germany}

\date{\today}

\begin{abstract} 
 
 Progress in simulating QCD at nonzero baryon density requires, amongst others, substantial numerical effort.
  Here we propose two different expansions to all orders in the hopping parameter,  preserving the full Yang-Mills action, which are much cheaper to 
simulate.
We carry out simulations using complex Langevin dynamics, both in the hopping expansions and in the full theory, for two flavours of Wilson fermions, and agreement is seen at sufficiently high order in the expansion.  These results provide support for the use of complex Langevin dynamics to study QCD at nonzero density, both in the full and the expanded theory, and for the convergence of the latter.
 
\end{abstract}


\maketitle


{\em Introduction --}
 Strongly interacting matter at finite density is a highly topical and interesting problem,
relevant for a wide spectrum of phenomena - from heavy ion collisions to neutron stars and the phases of QCD. The study of QCD at nonzero density (or quark chemical potential) is challenging, however, by the difficulty of performing ab initio lattice QCD calculations, due to what is known as the sign problem for numerical simulations; see e.g. \cite{Aarts:2013bla}. 
The sign problem appears in theories where the partition function is represented as a sum (or integral) over configurations with a complex associated weight factor. As such it is not unique to QCD and indeed the sign problem appears in many theories with an imbalance between particles and antiparticles \cite{Chandrasekharan:2008gp}. Moreover, the presence and severity of the sign problem depends 
on the representation of the path integral, which has allowed for  solutions in certain cases \cite{Gattringer:2014nxa}. 

In the case of QCD, until now only the small  $\mu/T$ region could be investigated \cite{deForcrand:2010ys}. In the past years, however, 
 progress in complex Langevin (CL) dynamics \cite{Aarts:2009uq,Seiler:2012wz,Sexty:2013ica}
 has led to the hope that the full region of physical interest can be explored. 
Nevertheless, this approach still implies demanding calculations, especially at low temperature, where large lattices are required. It is therefore useful to combine it with the hopping parameter expansion, 
 which can be formulated as a systematic approximation for QCD at finite chemical potential. The hopping expansion yields 
an analytic series which for not too small quark masses is expected to converge towards full QCD at sufficiently high order, but is still much easier to simulate than the complete theory. 
The leading-order (LO) and next-to-leading order (NLO) terms have been determined in the past 
using a loop expansion  \cite{Bender:1992gn,Blum:1995cb,Aarts:2002} and 
used, together with the full Yang-Mills action, to explore the phase diagram \cite{DePietri:2007ak} 
(see also Ref.\ \cite{ben}).
This work has been followed by calculations using a combination of the strong coupling expansion for the Yang-Mills 
action and hopping parameter expansion for the determinant, also with the aim to determine the phase diagram
\cite{Fromm:2011qi,Fromm:2012eb,Greensite:2013vza,Langelage:2013paa,Langelage:2014vpa}. 
However, it becomes notoriously difficult to go to higher order in the general case. 
In this paper we present two alternative ways to introduce higher-order corrections in the hopping parameter expansion.
They allow the calculation of corrections {\em to any order} preserving the
full Yang-Mills action and without having to consider explicitly the
 fermionic loops and their combinatorial factors. Hence this procedure 
improves the current status substantially and
aims directly at approaching full QCD at any coupling.
 The expansions have different merits: the
$\kappa$-expansion introduced below is numerically cheaper, but converges well only 
at not too high chemical potential, while the slightly more expensive
$\kappa_s$-expansion converges also at larger values.
They may also have better convergence properties than 
the loop expansion \cite{DePietri:2007ak}, where the effective expansion parameter is  
$N_\tau \kappa$, with $N_\tau$ the extent in the temporal direction.
 
In order to solve the resulting theories (which still suffer from a sign problem) numerically, we demonstrate that this can be achieved using CL dynamics. We compare the results with CL simulations of the {\em full} theory \cite{Sexty:2013ica}. We find excellent agreement at sufficiently high order in the expansion, providing justification for both the expanded and the full results. We suggest that the approach may 
provide access to the QCD phase diagram beyond what can be normally achieved  \cite{deForcrand:2010ys}.

{\em Hopping parameter expansions --}
 We consider lattice QCD with Wilson fermions at nonzero quark chemical potential $\mu$. 
After integrating out the fermions, the QCD partition function is given by
\be
Z = \int DU\, \e^{-S},\quad S= S_{\rm YM}- \log \det M 
\label{e.zqcd},
\ee
where $S_{\rm YM}$ is the standard Wilson gauge action and the fermion matrix (for each flavour) is given by
\be
M = 1-\kappa Q = 1-\kappa_s S - R,
\label{eqMQCD}
\ee
with
\bea
 S_{xy} &=& 2\sum_{i=1}^3  \left( \Gamma_{-i} U_{x,i} \delta_{y,x+a_i}  
+ \Gamma_{+i} U_{y,i}^{-1} \delta_{y,x-a_i}  \right), \\
 R_{xy} &=&  2\kappa\left( e^\mu \Gamma_{-4} U_{x,4} \delta_{y,x+a_4}  
+ e^{-\mu} \Gamma_{+4} U_{y,4}^{-1} \delta_{y,x-a_4}\right). \nn
\eea
We introduced a separate parameter $\kappa_s$ for hopping in the spatial direction, 
which will be used to define the 
$\kappa_s$-expansion below, with the identification $\kappa_s = \kappa$.
The hopping parameter is (naively) related to the bare quark mass $m$ as $\kappa=1/(2m+8)$. 
The projectors are $\Gamma_{\pm \nu} = (1 \pm \gamma_\nu)/2$, with $\Gamma_{\pm \nu}^2 = \Gamma_{\pm \nu}$ and $\Gamma_{+\nu}\Gamma_{ -\nu} =0$.
For nonzero $\mu$, $\det M$ is complex, but it satisfies the 
relation $[\det M(\mu)]^* = \det M(-\mu^*)$, due to $\gamma_5$-hermiticity of the Dirac matrix $M$.
As always, the temperature is given by the inverse temporal length of the lattice, $T=1/(aN_\tau)$
(we use lattice units $a\equiv 1$).

In a straightforward hopping expansion, to which we will refer to as the {\em $\kappa$-expansion}, $\det M$ is expanded as
\be
\label{eq:kappaexp}
 \det M = \det(1-\kappa Q) = \exp\sum\limits_{n=1}^{\infty}  
- {\kappa^n \over n} \tr\,Q^n.
\ee
For $N_f>1$ flavours their contributions are summed in the exponent.
Since $Q$ contains hoppings, only even powers contribute in the expansion, due to the traces. 
Fermionic observables are calculated using the same expansion: e.g.\ the chiral condensate and baryonic density are written as
 \bea
  \bra \bar \psi \psi \ket &=& \frac{2\kappa N_f}{\Omega} \label{meas-cc}
  \sum_{n=0}^\infty \kappa^n  \left\bra \tr\, Q^n \right\ket, \\
 \bra n\ket &=& -\frac{N_f}{\Omega} \sum_{n=1}^\infty \kappa^n \left\bra
 \tr \left({ \partial Q \over \partial \mu} Q^{n-1}\right) \right\ket, \label{meas-dens}
\eea
with $\Omega=N_s^3N_\tau$ the lattice volume.
A drawback of this approach is that one needs to go to order $N_\tau$ before $\mu$ dependence is visible, 
since $\mu$ dependence only arises when loops can wind around the lattice in the time direction. Moreover, the terms in the expansion contain $e^\mu$ contributions, which affects the convergence for large $\mu$.

For those reasons we consider a second scheme, an expansion in the spatial hopping parameter only, to which we refer as the {\em $\kappa_s$-expansion}.
At LO, obtained by taking $\kappa_s=0$ in Eq.\ (\ref{eqMQCD}), we recover heavy dense QCD (HDQCD), which formally relies \cite{Bender:1992gn} on the double limit
$\kappa \rightarrow 0$, $\mu \rightarrow \infty$, $\zeta \equiv 2\kappa\e^{\mu} \;\; {\rm fixed}$. 
Note that in this limit only Polyakov loops ${\cal P}_\xv$  survive;
inverse Polyakov loops ${\cal P}^{-1}_\xv$ are introduced to preserve the 
$\gamma_5$-hermiticity and allow a smooth continuation to small $\mu$ \cite{Aarts:2008rr}.  
The HDQCD determinant reads
\be
 \det M_{\rm LO} = \prod_\xv 
 \det \left(1+ C{\cal P}_\xv \right)^2 
 \det \left(1+ C' {\cal P}_\xv^{-1}\right)^2,
 \label{e.det0} 
\ee
where $C(\mu)=(2\kappa e^\mu)^{N_\tau}$ and $C'(\mu)=C(-\mu)$. The remaining determinants in colour space are easily expressed in terms of traced (conjugate) Polyakov loops,  
$P_x=\tr\, {\cal P}_\xv/3$, $P_x'=\tr\, {\cal P}^{-1}_\xv/3$ \cite{Aarts:2008rr,DePietri:2007ak}. Since quarks cannot hop in space, this corresponds to the static limit.

HDQCD can be used as an approximation to QCD for any $\kappa$ and $\mu$. However, since the hopping expansions are analytic, one can go further and consider successive higher order terms in $\kappa_s^2$  in the loop expansion of the determinant, by using decorated 
Polyakov loops \cite{DePietri:2007ak}.
 To go beyond LO systematically in the $\kappa_s$ expansion, we separate the temporal part and write
\be
M = (1-R) \left( 1 - \frac{1}{1-R} \kappa_s  S \right).
\ee
 The full determinant is then expanded as
\be
\det M = \det (1-R) \exp \sum\limits_{n=1}^{\infty}  
- \frac{\kappa_s^n}{n} \tr\left(  \frac{1}{1-R}  S \right)^n.
 \label{kappasdet}
\ee
Again, only even powers have a nonvanishing contribution in the expansion.
The determinant and  inverse of $1-R$ can be calculated analytically.
 Since (unlike for staggered fermions) backtracking is forbidden, the former is 
just given by the LO expression (\ref{e.det0}). 
For the inverse we  write $(1-R)^{-1} = (1-R^+)^{-1} + (1-R^-)^{-1} -1$ 
with $R^{\pm }$ containing the timelike hoppings in the positive/negative 
direction. We then expand
\be
(1-R^+)^{-1}_{xy}= \sum_{n=0}^{\infty}\left(2\kappa e^\mu \Gamma_{-4} 
U_{x,4} \delta_{y,x+a_4}\right)^n,
\label{eq:10}
\ee
 and similarly for $(1-R^-)^{-1}$. The geometric series
 can easily be resummed introducing temporal strings between
  $x$ and $y$. 
  The fermionic observables can be expressed similarly to Eqs.\ (\ref{meas-cc},\ref{meas-dens}). 
 A similar expansion was used in Refs.\ \cite{Langelage:2013paa,Langelage:2014vpa}  in combination with the strong 
coupling expansion for the Yang-Mills action to develop an effective Polyakov-loop 
model for QCD. These papers also 
contain detailed formulae for the relations following
from Eqs.\ (\ref{kappasdet}, \ref{eq:10}).
Here our main interest is to develop a model-free method for QCD, using 
the full Yang-Mills action and the hopping parameter expansion to arbitrary 
order, allowing calculations in a large region of parameters.

{\em Complex Langevin simulations --}
 We define the N$^q$LO approximation by truncating the expansions (\ref{eq:kappaexp}, \ref{kappasdet}) to order $\kappa_{(s)}^{2q}$. The resulting Boltzmann  weights are however still complex, invalidating importance sampling; hence we use CL instead.  
 Since CL does not rely on positivity of the weight, it has the potential
to simulate lattice models for which importance sampling fails.
The approach \cite{parisi,klauder} is based on setting 
up a stochastic process on the {\it complexification} of the configuration space. 
One can formally prove \cite{Aarts:2009uq} correctness of the approach, provided that
certain conditions are met, such as a rapid decay of the probability distribution effectively sampled on the complexified configuration space and holomorphy of the drift term and observables. 
Numerical problems, e.g.\  runaway trajectories, may be present due to the amplification of 
unstable modes in the drift dynamics by numerical imprecision, but these can to a large extent be taken care of by adaptive step sizes \cite{Aarts:2009dg}. Generally, one has significant freedom in defining the process for a given action \cite{Aarts:2012ft}.

In lattice QCD links live in SU(3) and the Langevin equation reads \cite{batrouni}
\bea
U_{x,\nu}\mapsto \exp\left\{\sum_a i\lambda_a (\epsilon 
K_{x\nu a}+\sqrt{\epsilon}\eta_{x\nu a})\right\}U_{x,\nu}, \;\;
\label{dyn}
\eea
where $K_{x\nu a}= -D_{x\nu a}S $ is the drift force, $\epsilon$ the (adaptive) stepsize, and $\eta$ independent Gaussian noises satisfying
$\bra \eta_{x\nu a} \eta_{x'\nu'a'}\ket =2 \delta_{aa'} \delta_{xx'} \delta_{\nu\nu'}$.
 A complex action leads to a complex drift force $K$, and links take values in SL(3,$\mathbb{C}$).
 While the drift term is gauge covariant and transverse to the gauge orbits, the noise term contains components along the gauge orbits. Solutions of the stochastic process may therefore go far from the unitary submanifold, resulting in a wide distribution in the noncompact direction, undermining one of the conditions for the validity of the approach.
To deal with this problem, we use the method of gauge cooling \cite{Seiler:2012wz} (see also the review \cite{Aarts:2013uxa}), which uses noncompact gauge transformations to force the process to stay near the unitary manifold. 
This leads to a thin distribution 
which is required for the convergence proof
and together with the adaptive stepsize this practically eliminates runaways.

Another complication is due to zeroes in the measure, i.e.\ $\det
M=0$, leading to a meromorphic drift. Poles in the drift may provoke
wrong convergence of the process, as shown in nontrivial, soluble
models \cite{Mollgaard:2013qra}.  At any finite order $n$ the
$\kappa$-expansion for the drift has no poles and the possible zeroes
in the determinant will only show up as poor convergence of the series
when approaching such configurations.  Its drawback at large $\mu$ comes
from the $e^\mu$ contributions to the terms in the expansion. The
$\kappa_s$-expansion, which takes care of these
contributions analytically, does not have this drawback, but here the drift itself
will have singularities at the zeroes of the LO determinant,
det$(1-R)$. These singularities may cause problems in some regions of
parameter space. While a systematic understanding of the effect of poles
is still missing, the results presented below indicate that it is 
not an issue here.

HDQCD has been first 
studied using CL in Ref.\ \cite{Aarts:2008rr}. Supplemented with the gauge cooling procedure,  
it produces correct results (by comparison with reweighting data where
available) \cite{Seiler:2012wz}, provided that the gauge coupling  $\beta\gtrsim 5.7$. Below this threshold, gauge cooling is not effective enough to control the distribution in SL(3, $\mathbb{C}$). The threshold depends only very mildly on the lattice size or $\mu$ \cite{Aarts:2013nja}. Hence, the continuum limit can be reached by increasing $\beta$.
The extension to full QCD was given in Ref.~\cite{Sexty:2013ica} for staggered quarks. Since the
gauge transformations are interlaced with dynamical steps, the
numerical costs are increased by a factor without changing the volume dependence of the algorithm.

The expressions for the Langevin drift at LO can be found in Ref.\ \cite{Aarts:2008rr}. 
 To go beyond LO, we now discuss how the drift terms using the  
$\kappa$- and $\kappa_s$-expansions are implemented in CL dynamics.
In the case of the $\kappa$-expansion, the contribution to the drift (for a single flavour) is
\bea
K_{x\nu a}&=&  -\sum_{n=1}^\infty \kappa^n\tr \left( Q^{n-1} 
D_{x\nu a} Q \right).
\eea
In the $\kappa_s$-expansion, there are two terms, with the contribution from the first factor in Eq.\ (\ref{kappasdet}) given by the drift at LO \cite{Aarts:2008rr}. The contribution from the second factor is  
\bea
K_{xia} &=&  - \sum\limits_{n=1}^\infty  \kappa_s^n 
\tr \left({1\over 1-R} (D_{xia} S)  
 \left[ {1 \over 1-R}  S \right]^{n-1} \right), \nn \\
K_{x4a} &=&  - \sum\limits_{n=1}^\infty \kappa_s^n 
\tr \left( {1 \over 1-R } (D_{x4a} R)  
 \left[  {1 \over 1-R}  S \right]^{n} \right), \;\;\;\; 
\eea
for spatial and temporal links, correspondingly.
In the numerical implementation the traces are computed using a 
noisy estimator, i.e.\ by choosing a Gaussian random vector $ \eta_i$  
(here $i$ represents space-time, colour and Dirac indices) satisfying
$\bra \eta_i \ket=0$, $\bra \eta^*_i \eta_j \ket= \delta_{ij}$, and 
then constructing the drift as 
\bea
K_{x\nu a}= \bra\eta^* (D_{x\nu a} Q)  s\ket, \quad\quad
  s = -\sum_n \kappa^n Q^{n-1} \eta, \quad
\eea
for the $\kappa$-expansion.
The dominant numerical cost of the fermions, when including corrections up to $\kappa^n$, 
is thus $n-1$ multiplications with the sparse matrix $Q$.
 For the $\kappa_s$-expansion the drift is computed in a similar fashion, but the numerical cost is slightly higher, as additional multiplications with the matrix $(1-R)^{-1}$ are required.
 We emphasise that numerical inversion of the fermion matrix $M$ is not required at any stage.

\begin{figure}[h]
\begin{center}
\includegraphics[width=0.98\columnwidth]{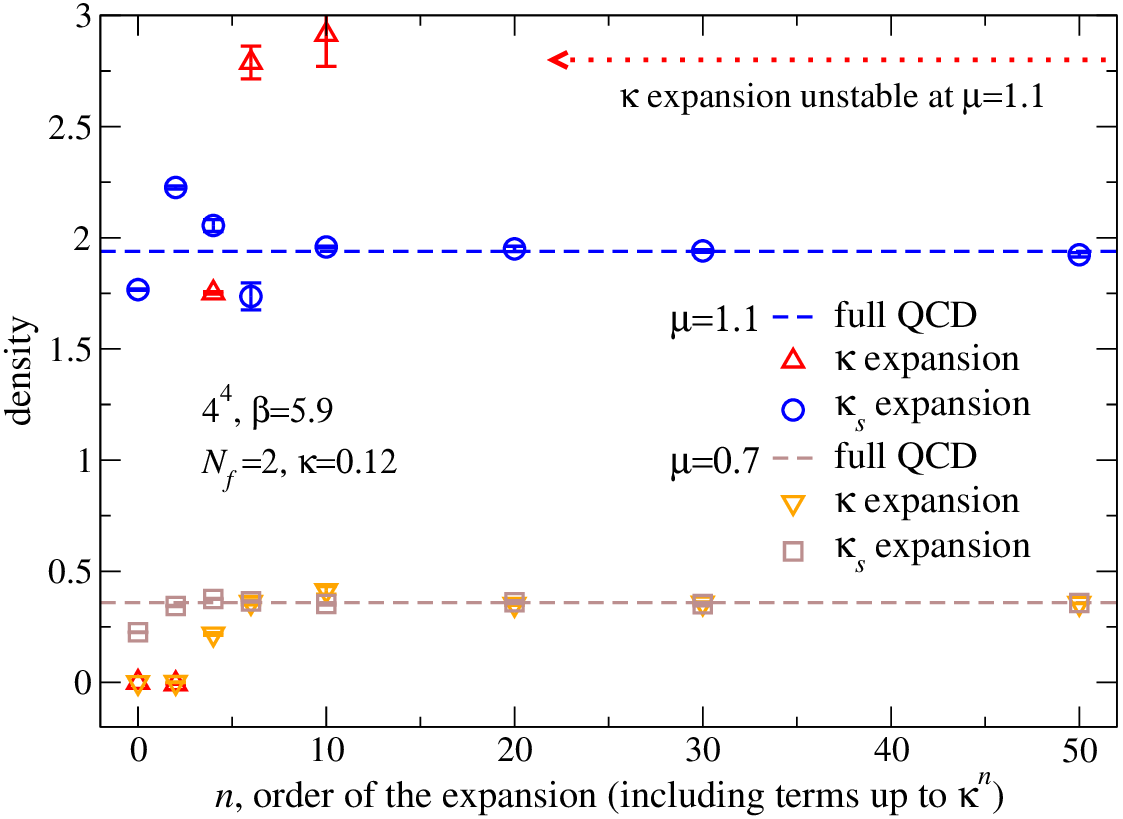}  
\caption{Dependence of the quark density (in lattice units) on the order of the truncation of the $\kappa$- and $\kappa_s$-expansions, for $\mu=0.7$ and $1.1$, on a $4^4$ lattice with $\beta=5.9$, $\kappa=0.12$, and $N_f=2$. The region where the $\kappa$-expansion breaks down for $\mu=1.1$ is indicated.
The lines show the result for full QCD.
Saturation density is $n_{\rm sat}=2N_cN_f=12$.
}
\label{f_nlopp_dens}
\end{center}
\end{figure}

{\em Simulation results --}
 In order to demonstrate the feasibility of the approach we have carried out simulations for $N_f=2$ flavours. Here we present first results on a $4^4$ lattice, at $\beta=5.9$ and $\kappa=\kappa_s=0.12$, for several values of $\mu$. We always use lattice units. 
We also compare with results for full QCD, obtained with complex Langevin dynamics, extending the approach of Ref.\ \cite{Sexty:2013ica} for staggered fermions to Wilson fermions.
We have measured the scale using the gradient flow, 
as proposed in Ref.\ \cite{Borsanyi:2012zs}. The scale depends on the theory that is considered: for HDQCD we find that $\beta=5.9$ and $\kappa=0.12$ corresponds to  $a \simeq 0.12$ fm, while for full QCD we find $a \simeq 0.114$ fm.

\begin{figure}[t]
\begin{center}
\includegraphics[width=0.98\columnwidth]{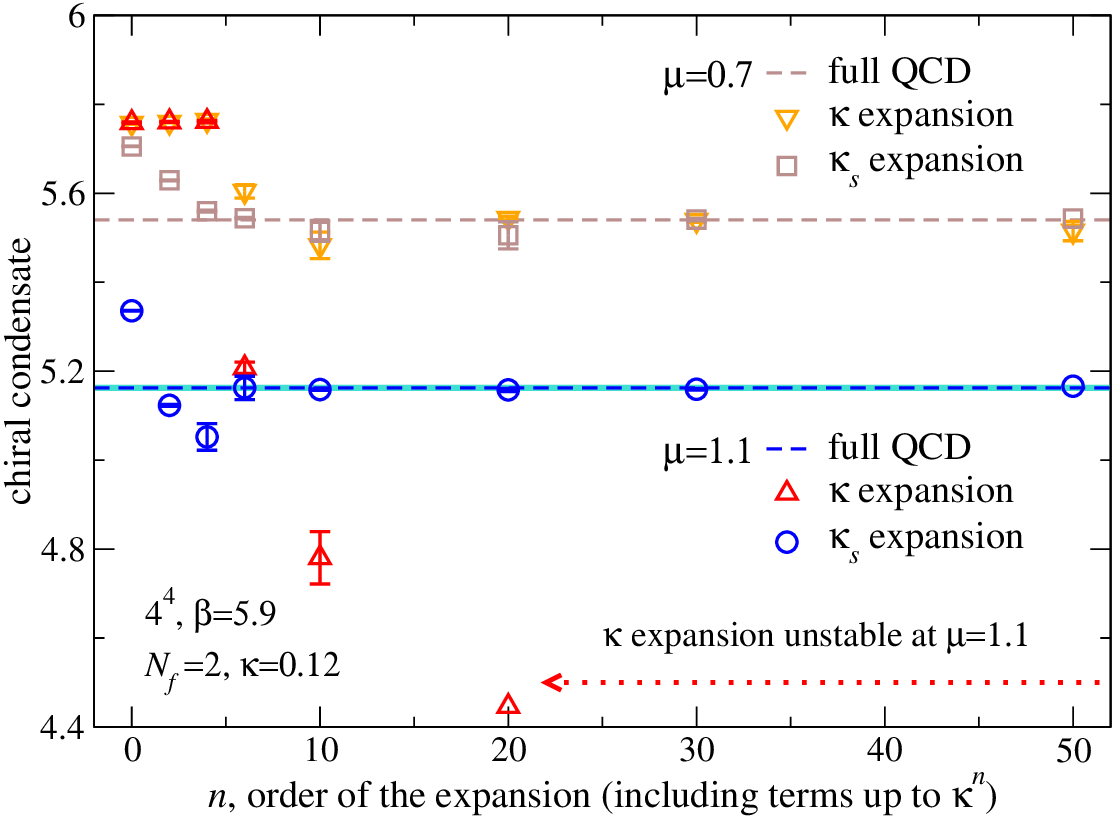}  
\includegraphics[width=0.99\columnwidth]{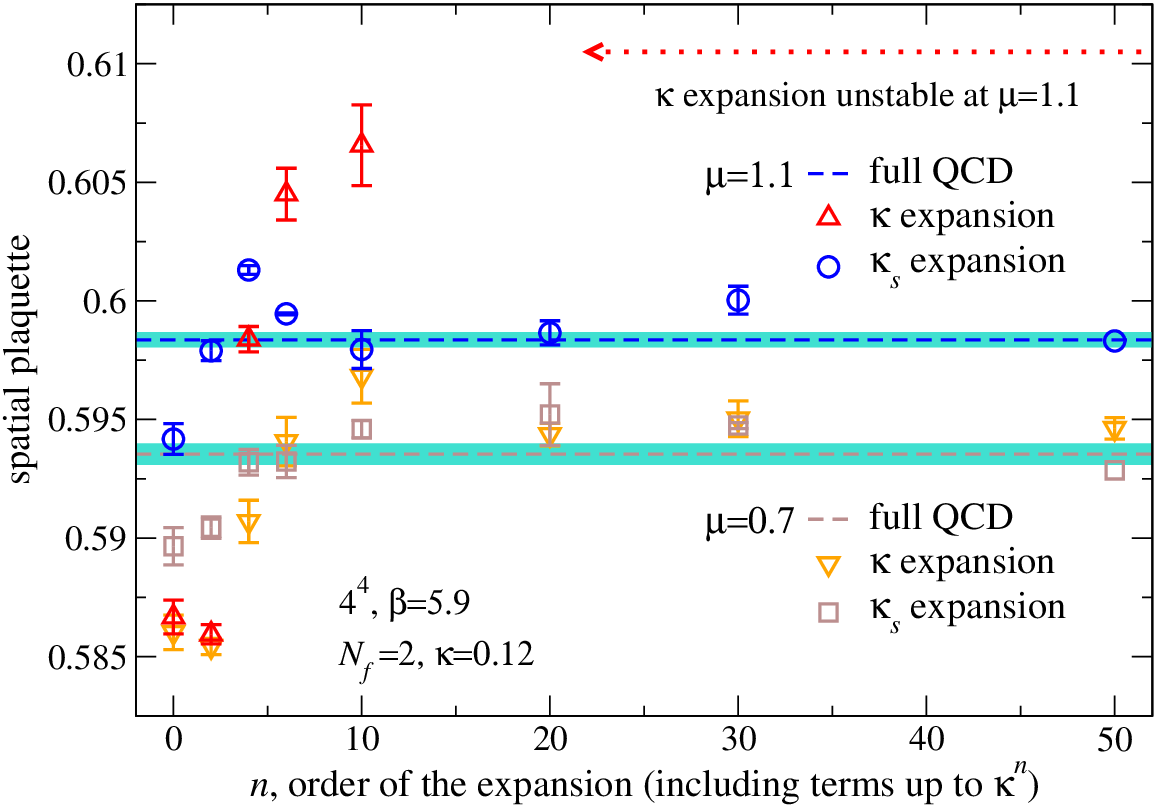}  
 \caption{As in Fig.\ \ref{f_nlopp_dens}, for the chiral condensate (top) and the spatial plaquette (bottom).
 }
\label{kks_plaq}
\end{center}
\end{figure}

In Fig.~\ref{f_nlopp_dens} we present the quark number density as a function of the order in the $\kappa$- and $\kappa_s$-expansions for two different $\mu$ values. We also show the result for full QCD.
At the smaller $\mu=0.7$ (or $\mu/T=2.8$) the two expansions perform similarly, but at the larger $\mu=1.1$ (or $\mu/T=4.4$) the $\kappa$-expansion breaks down, due to the presence of $\kappa e^\mu$ terms in the series.
The $\kappa_s$-expansion converges in both cases (already at order $\kappa^{10}$), as the $\mu$ dependence does not interfere with the spatial hopping expansion.  In the $\kappa$-expansion, one needs to go to order $\kappa^4$ to find the first $\kappa$ dependence, since only then it can appear in a closed loop, i.e.\ the plaquette. Similarly, one needs to go to order $\kappa^{N_\tau}$ to find the first $\mu$ dependence. Hence the results up to $\kappa^2$ equal the quenched  ($\kappa=0, \mu=0$) result.  In the $\kappa_s$-expansion, there is $\kappa$ and $\mu$ dependence at all orders. We not only observe good convergence, but also agreement with the full result. This is a nontrivial test for both expansions and for the CL simulation of the full theory, in particular it implies that the lack of holomorphicity due to poles in the drift arising from the determinant does not invalidate the results.
Similar behaviour is observed for the chiral condensate and the spatial plaquette, see Fig.~\ref{kks_plaq}. 

\begin{figure}[t]
\begin{center}
\includegraphics[width=0.98\columnwidth]{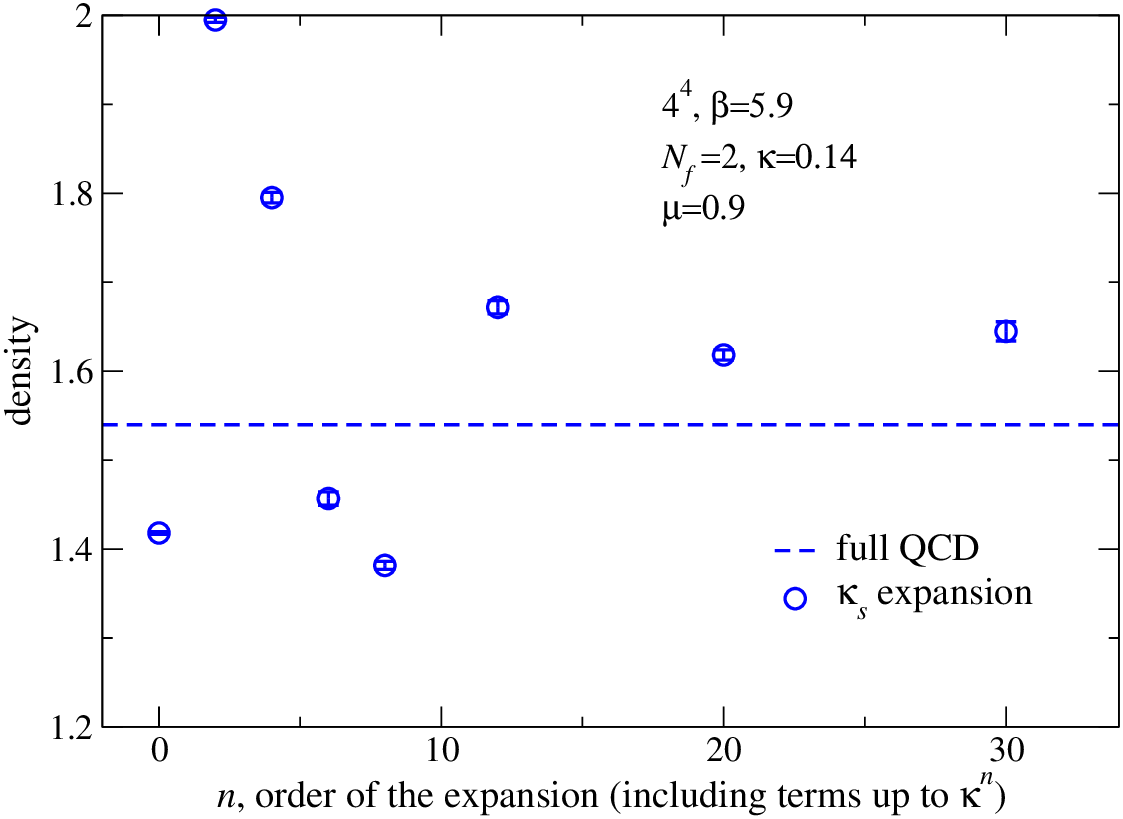}  
\includegraphics[width=0.98\columnwidth]{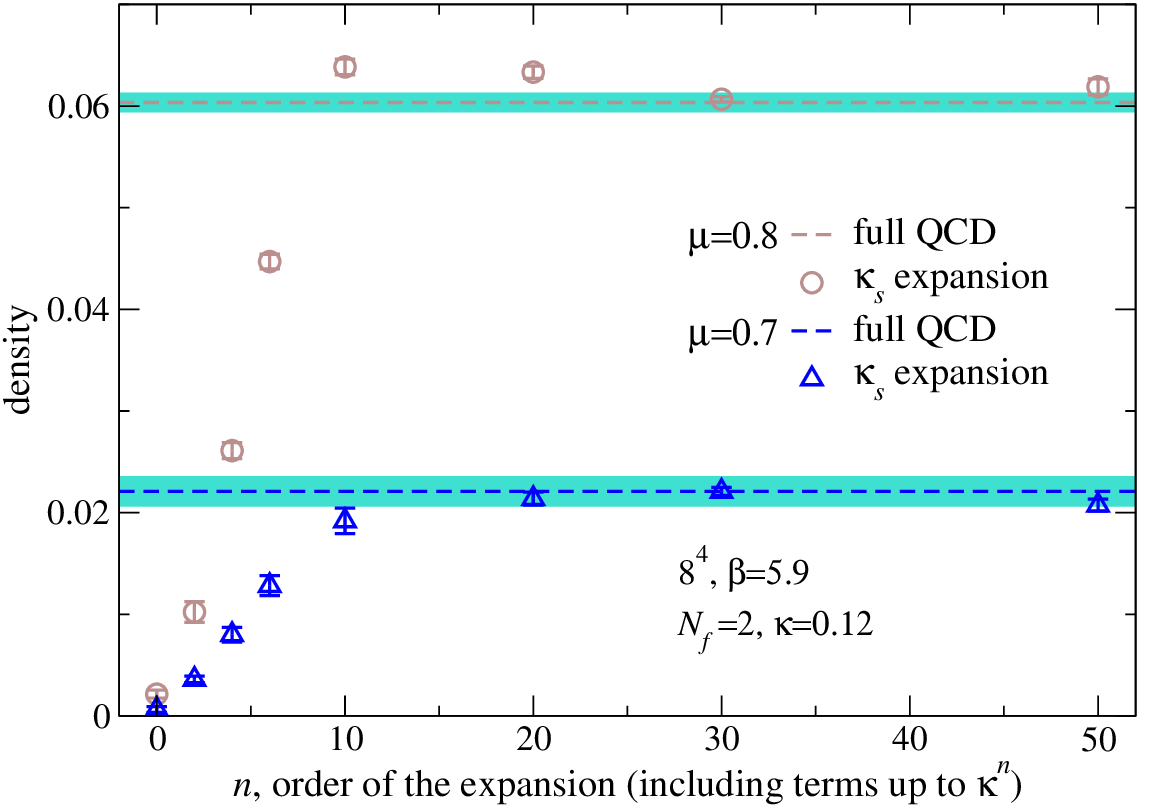}  
 \caption{As in Fig.\ \ref{f_nlopp_dens}, density at $\kappa=0.14$ on a $4^4$ lattice (top) and at $\kappa=0.12$ on a $8^4$ lattice (bottom), in both cases for the $\kappa_s$-expansion and full QCD.
  }
\label{fig:8conv}
\end{center}
\end{figure}

The $\kappa_s$-expansion appears to converge relatively quickly for $\kappa=0.12$, at orders where it is still much cheaper to simulate than full QCD. The convergence radius depends on the parameters used and the expansion appears to break down below $ \kappa=0.14$ at $\mu=0.9$, see Fig.~\ref{fig:8conv} (top). 
 However,  first results indicate that the convergence radius 
does not seem to depend on the lattice size. Fig.~\ref{fig:8conv} (bottom) shows the convergence of the expansion
on a larger  $8^4$ lattice, corresponding to a lower temperature $\sim 200$ MeV, for two $\mu$ values, corresponding to $\mu/T=5.6$ and $6.4$.  Here we note that at $\mu=0$ the theory is in the confined phase. 
The rapid rise of the density as $\mu$ increases from 0.7 to 0.8 indicates that at the larger $\mu$ value the theory is no longer confining. This interpretation is indeed  supported by a nonzero Polyakov loop expectation value (not shown).
A detailed study of how the convergence depends on the chemical potential, hopping parameter and lattice volume requires further investigation.

{\em Summary --}
 We have outlined a programme to simulate QCD at nonzero density using two distinct hopping parameter expansions and demonstrated that the approach appears feasible, at least for not too small (bare) quark masses.
We used two formulations: the $\kappa$- and $\kappa_s$-expansions, which can be effectively and cheaply calculated, using complex Langevin dynamics, to high order. Hence convergence can be checked explicitly.
We have also presented for the first time complex Langevin simulations of full QCD with Wilson fermions at finite chemical potential. This  makes it possible to compare the hopping parameter expansion results with those obtained directly in full QCD. The agreement in this case can also be used to justify the full Langevin results, and, in particular, to demonstrate that the nonholomorphicity of the action due to the determinant does not invalidate the Langevin results. 
We emphasize  that the approach outlined here is free of further approximations and based solely on the hopping parameter expansion to high order. Hence model-dependent features of effective models, such as those obtained using the strong coupling expansion, can be verified with our approach as well as by comparing with full QCD results.

While the first results are encouraging, there are many aspects that require further investigation, in particular the behaviour on larger lattices at lower temperature and large ${\mu}/{T}$, 
as well the convergence properties when decreasing the quark mass (increasing the hopping parameter). 
First results indicate that there is no problem in reducing the temperature and hence we propose to use the methods described here to investigate the phase diagram of QCD 
in regions not accessible otherwise, using larger lattices at sufficiently high order in the $\kappa_s$-expansion. We hope to come back to this in the near future.

\acknowledgments

We thank Benjamin Jaeger and Felipe Attanasio for collaboration on related topics, and 
 BMBF and MWFK Baden-W\"urttemberg  (bwGRiD cluster) for support.
ES and IOS are supported by Deutsche Forschungsgemeinschaft.
GA is supported by STFC, the Royal Society, the Wolfson Foundation and the Leverhulme Trust.

\end{document}